\documentclass[a4paper,11pt]{article}
\usepackage{aaskaiid}
\usepackage{orcidlink}

\usepackage{soul}

\title{Low-frequency VLBI with the SKA-Low}
\ShortTitle{VLBI with SKA-Low}

\author[1,2]{R.~Timmerman\orcidlink{0000-0001-9404-2612}}
\ShortName{Timmerman et al.} 
\author[3]{E.~L.~Escott\orcidlink{0009-0009-5108-4324}}
\author[3,4]{T.~Vernstrom\orcidlink{0000-0001-7093-3875}}
\author[1,2]{L.~K.~Morabito\orcidlink{0000-0003-0487-6651}}
\author[3]{C.~Reynolds\orcidlink{0000-0002-8978-0626}}

\affiliation[1]{Centre for Extragalactic Astronomy, Department of Physics, Durham University, Durham DH1 3LE, UK}
\emailAdd{roland.timmerman@durham.ac.uk}
\affiliation[2]{Institute for Computational Cosmology, Department of Physics, Durham University, South Road, Durham DH1 3LE, UK}
\affiliation[3]{Australia Telescope National Facility, CSIRO, Space and Astronomy, PO Box 1130, Bentley, WA 6102, Australia}
\affiliation[4]{International Centre for Radio Astronomy Research (ICRAR), The University of Western Australia, 35 Stirling Highway}

\abstract{The Square Kilometre Array Observatory (SKAO) will provide unprecedented observational capabilities, revolutionizing radio astronomy in the next decade. Of particular interest for many science areas is the low-frequency component: the SKA-Low. This array, operating at frequencies between 50 and 350~MHz, will be able to observe the Southern hemisphere with an angular resolution of several arcseconds. However, many science cases demand finer angular resolutions than the current design baseline for the SKA-Low. In this chapter, we present the Low-frequency Australian Megametre-Baseline Demonstrator Array (LAMBDA) project: the construction of a several SKA-Low-like stations throughout Australia which can be employed to extend the SKA-Low's baselines to the scale of megameters. This allows LAMBDA to not only probe the scientific opportunities accessible at higher angular resolutions, but also prove the feasibility of a potential extension of the SKA-Low to such baselines. Finally, LAMBDA also serves to characterize the performance of the SKA-Low as part of a Very Long Baseline Interferometry (VLBI) network and by providing an early testbed for the calibration strategies which would be required. With LAMBDA, the SKA-Low will be able to make a major impact in many science areas, including but not limited to exoplanets, resolved active galactic nuclei (AGN), young stellar objects and supernova remnants, pulsar astrometry and scintillometry, gravitational lensing and fast radio bursts (FRBs).}

\begin{document}
\maketitle

\section{Introduction}

The Square Kilometre Array Observatory (SKAO) is set to revolutionize radio astronomy in the 2030's with its two sets of telescopes: SKA-Low in Australia and SKA-Mid in South Africa. Although both of these arrays will offer groundbreaking sensitivity in their respective frequency bands, they will do so at strongly varying angular resolutions. According to the design baseline, the SKA-Mid telescopes, operating between 350~MHz and 15.4~GHz, will feature a maximum baseline length of 159.6~km \citep{braun2019anticipatedperformancesquarekilometre}. At a reference frequency of 1.5~GHz, this will result in an angular resolution of a quarter of an arcsecond. Meanwhile, the SKA-Low telescopes, despite observing at frequencies between 50 and 350~MHz, will only have a maximum baseline length of 73.4~km. At a reference frequency of 150~MHz, this will provide an angular resolution of around 6 arcseconds. Such an angular resolution is insufficient to match the SKA-Mid for broadband spectral index mapping. Additionally, we know that the vast majority of radio sources will remain unresolved at this angular resolution \citep{shimwell22}, proving that many science cases require substantially finer angular resolutions at low radio frequencies.

As has been shown for the past decade in the Northern hemisphere with the Low Frequency Array \citep[LOFAR;][]{haarlem13}, very long baseline interferometry (VLBI) at low frequencies is opening up exciting new science \citep{morabito25}. While technically challenging, high resolution studies at low and ultra-low frequencies is one of the least explored parameter spaces. There currently exist many facility or baseline options for VLBI observing with SKA-Mid, however, current options for long baselines at the SKA-Low frequencies are very limited. 

To provide a pathway to high angular resolution observations with the SKA-Low, in addition to investigating possible international baseline partners, the Commonwealth Scientific and Industrial Research Organisation (CSIRO) in Australia is developing the Low-frequency Australian Megametre-Baseline Demonstrator Array (LAMBDA). LAMBDA is an array comprised of several SKA-Low-like stations which can be employed as an extension to SKA-Low and produce megametre-scale baselines. With LAMBDA, we not only demonstrate the feasibility of extending SKA-Low into the continent-scale array it was originally envisioned to be, but we also provide motivation for a future expansion of SKA-Low by probing the scientific value provided by such observational capabilities.

\section{Possible VLBI extensions to SKA-Low}
A number of international facilities have the potential to participate in SKA-Low VLBI observations. Here we give an overview of low-frequency opportunities in countries with shared skies with SKA-Low. Parts of this section build upon discussions and material developed within the SKA-Low VLBI community, in particular outcomes of discussions at the SKA-Low VLBI community meeting held in 2025.

\subsection{India}

India hosts two facilities well suited for participation in low-frequency VLBI in the SKA era: the upgraded Giant Metrewave Radio Telescope \citep[uGMRT;][]{swarup91,gupta17}, located in western India, and the Ooty Radio Telescope \citep[ORT;][]{swarup71,subrahmanya17}, situated in the south of the country. The uGMRT operates over a wide frequency range from 110 to 1450~MHz with nearly continuous frequency coverage, while the ORT operates at 325~MHz with a current bandwidth of 16~MHz, soon to be expanded to 40~MHz. The baseline between uGMRT and ORT is approximately 900~km, and both would form baselines of about 8000~km with SKA-Low. Low-frequency VLBI capability has already been demonstrated in India through a combination of uGMRT and ORT observations, including multi-station experiments at P-band and subsequent integration into international VLBI efforts. These include uGMRT three-station VLBI tests at P-band (325~MHz) with Iitate Observatory in Japan \citep{iwai12}, while uGMRT has also conducted successful L-band VLBI experiments with stations in Australia, Japan, and the European VLBI Network (EVN).

The uGMRT currently supports the formation of four independently steerable tied-array beams within the primary beam of each GMRT antenna, offering valuable flexibility for certain types of VLBI observations. Both uGMRT and ORT employ in-house recording systems built from commercial off-the-shelf (COTS) components that generate data in native formats, which are subsequently converted to two-thread VLBI Data Interchange Format (VDIF) through offline signal-processing algorithms. Efforts are underway to integrate the DiFX software correlator into their systems to enable in-house VLBI data correlation and analysis. Both facilities are prepared for expanded VLBI testing and collaborative experiments at P-band, and potentially at L-band, with other suitably equipped observatories.

\subsection{China}

Chinese researchers possess extensive expertise in VLBI through their participation in the Chinese VLBI Network (CVN) and the East Asian VLBI Network \citep[EAVN;][]{akiyama22}. Several facilities in China have potential to serve as SKA-Low VLBI partner stations, including the 21 Centimeter Array \citep[21 CMA;][]{zheng16}, comprising 127 antennas distributed across 40 stations, and the Five-hundred-meter Aperture Spherical Telescope \citep[FAST;][]{nan11}, a 110~m single-dish instrument. Additional candidates include the Tianlai array \citep{li20}, which consists of three adjacent north–south–oriented cylindrical reflectors, each measuring 15~m in length and 40~m in width, together with a pathfinder array of sixteen 6~m dishes. Also, the Interplanetary Scintillation (IPS) telescope array in Inner Mongolia \citep{yan18}, composed of three phased cylinders of 40~m × 120~m, has expressed interest in participating in SKA-Low VLBI and has initiated fundraising efforts to support modifications for VLBI operations.

\subsection{Japan}

Japanese researchers have a long and distinguished record in VLBI, having led and contributed to major initiatives such as the Haruka space-VLBI mission \citep{kobayashi00}, VLBI Exploration of Radio Astronomy \citep[VERA;][]{kobayashi03}, KaVA \citep{niinuma14}, the Japanese VLBI Network \citep[JVN;][]{doi16}, and the EAVN. The Iitate telescope operates at several frequencies relevant to SKA-Low, including 150, 220, and 320~MHz, and has been used in ongoing fringe tests with the uGMRT, ORT, and the Murchison Widefield Array \citep[MWA;][]{tingay13}. Successful fringe detections have been achieved between Iitate, uGMRT, and Ooty at 320~MHz, while a subsequent VLBI experiment between Iitate and MWA was conducted but affected by strong radio-frequency interference (RFI) at Iitate. Development of a VLBI recording system for SKA applications has commenced in Japan, utilising a PC-based architecture built from commercial off-the-shelf components. In addition, a BURSTT \citep{lin22} outrigger station located on Ogasawara Island extends coverage to frequencies as low as approximately 250~MHz, enabling potential VLBI observations with SKA-Low, though the sensitivity of this aperture array is limited at low elevations.

\subsection{South Korea}

Researchers in South Korea have extensive experiences of VLBI such as the Korean VLBI Netowrk \citep[KVN;][]{lee14}, KaVA, and EAVN. The Korea Astronomy and Space Science Institute (KASI) is planning to develop a GPU correlator, which is an upgrade version of current KVN GPU correlator, for SKA-VLBI. Additionally, a recent successful proposal for funding led by the Korea Advanced Institute of Science \& Technology (KAIST) will see the deployment of an SKA-Low like station (LAMBDA station, see Sect.~\ref{sect:lambda}) in South Korea. With this proposal will also be a joint research centre for low-frequency VLBI established between KAIST and ATNF in Australia.  

\subsection{Australia}

All of the possible baselines mentioned so far are located in the Northern hemisphere and consist of very long (\(>\)6000~km) mainly in a North-South direction with SKA-Low. This would provide long but isolated baselines, creating a poor \textit{uv}-coverage. Currently in Australia, the MWA also exists as a possible baseline to the SKA-Low. This would be a relatively short baseline of approximately 15-20~km. While the size of the MWA would provide good sensitivity, the long-term future of the MWA facility, e.g. past cycle 0 or 1 of SKA-Low, is still being decided.

For improved imaging fidelity, long baselines in the East-West direction and in the shorter to mid length range are required. Such baselines are not available through any existing observatories. Therefore, the construction of new stations spread across Australia forms the only way to obtain a contiguous and circular \textit{uv}-coverage.

\section{The LAMBDA project}\label{sect:lambda}

The Low-frequency Australian Megametre-Baseline Demonstrator Array is planned to consist of 4-6~stations distributed throughout Australia to produce megametre-scale baselines at the observing frequencies of SKA-Low. These stations will be able to operate as an extension to SKA-Low, but also independently for single-station use cases. As an extension of the SKA-Low, the LAMBDA stations will probe the scientific potential accessible at higher angular resolutions than the SKA-Low is set to provide, and provide validation of VLBI with SKA-Low.

Each LAMBDA station is planned to be located at a current site of the Long Baseline Array (LBA) in Australia. This ideally makes use of existing infrastructure for cost efficiency in relatively low radio-noise environments, with varying separations between stations. With these stations, LAMBDA will be able to reach baselines of up to $\sim$4000~km in length, offering an angular resolution better than 0.1 arcseconds at an observing frequency of 150~MHz, which closely matches the capabilities of the SKA-Mid. Importantly, this array configuration also provides sufficient intermediate baselines to provide reasonable image fidelity. Furthermore, LAMBDA also supports a further extension via the inclusion of the uGMRT in India, FAST in China or other low frequency telescopes in Asia to expand the horizons of VLBI with the SKA-Low to unprecedented scales. 

The LAMBDA stations will use the same antennas and layout as SKA-Low stations. Using the same dipole design as the SKA-Low provides a flexible, robust and well-tested design which will integrate smoothly with the SKA-Low. Furthermore, the overall station layout will match the SKA-Low in scale to provide the required sensitivity for calibration. To ensure flexibility and scalability during the development of LAMBDA, each station will be equipped with a custom backend.

\begin{figure}[h]
    \centering
	\includegraphics[width=0.7\columnwidth]{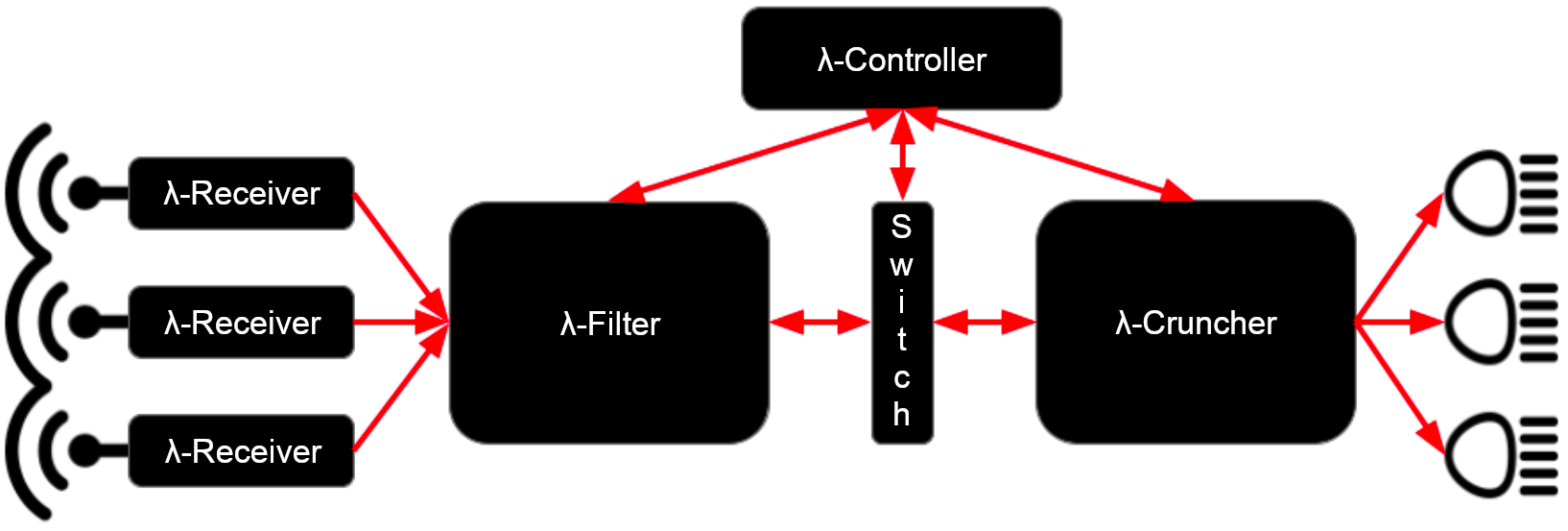}
    \caption{Schematic diagram of the LAMBDA backend following the signal path from left to right, showing the main components: the receivers, filter, switch and cruncher \citep[adapted from][]{reynolds24}.}
    \label{fig:receiver}
\end{figure}

The LAMBDA backend is designed to be highly flexible in order to both be able to form an extension to the SKA-Low for VLBI purposes, but also for a wide variety of other use cases (see the schematic in Figure~\ref{fig:receiver}). Each station will be equipped with a receiver based on CSIRO's BlueRing design \citep{hampson2019}, which is a custom module using commodity Radio Frequency System on Chips (RFSoCs). These receivers perform the digitization of the signal and provide these to the LAMBDA filter. This filter applies the required delay correction and runs the digitized data through a filterbank. Finally, it packetizes the data and appends timestamps. Optionally, the LAMBDA filter can form a single VLBI beam from all the station elements. The data is transmitted from the filter to a switch, which provides access to voltage data from all elements in a station to the LAMBDA cruncher. This cruncher applies the antenna calibration, performs the VLBI band reconstruction and converts to VDIF format for transfer to the correlator. Additionally, the cruncher will be able to perform immediate processing for various use cases, including pulsar timing, RFI excision, technosignature searches, custom filterbanks and the forming of additional beams.

\begin{figure}[h]
    \centering
	\includegraphics[width=0.7\columnwidth]{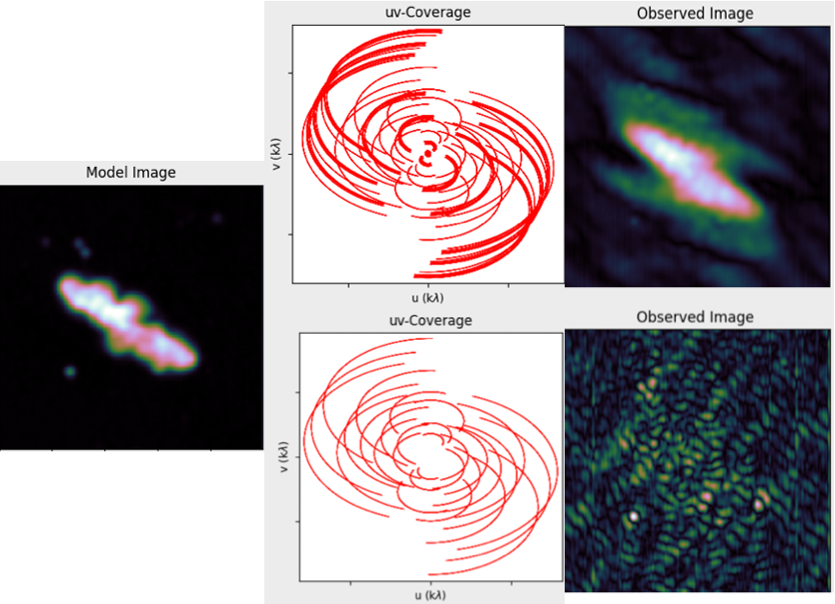}
    \caption{Illustration of how the image fidelity depends on LAMBDA's \textit{uv}-coverage in combination with the SKA-Low. The left-most panel shows the model image. In the upper row, the SKA-Low divided up into four sub-arrays, while in the bottom row, the SKA-Low is included as a single station. In the middle column, the \textit{uv}-coverage in both configurations is shown, with the resulting image shown on the right. We note no weighting is taken into account in these simulations.}
    \label{fig:uv}
\end{figure}

Although LAMBDA will provide long baselines on a range of physical scales, the array will strongly benefit from baselines on short to intermediate scales for any target that is substantially larger than the angular resolution of LAMBDA. For this purpose, it is valuable that the SKA-Low can be configured into four to eight separate sub-arrays (or VLBI beams). In Figure~\ref{fig:uv}, we compare the image quality obtained by including the SKA-Low as a single station versus as four sub-arrays in the LAMBDA array. As is clear from the figure, the recovery of diffuse emission requires the short baselines that SKA-Low can provide internally. Additionally, splitting up the array into separate sub-arrays also helps with the centrally-dominant sensitivity of the array.

\section{Scientific motivation}

As the only instrument offering high angular resolutions at low radio frequencies in the Southern hemisphere, LAMBDA has unique scientific potential.
Radiative processes such as synchrotron emission and electron-cyclotron maser instabilities typically result in spectra that are primarily or exclusively bright at low radio frequencies. Several science cases strongly depend on high angular resolutions radio observations in the low frequency regime, and LAMBDA's presence in the Southern hemisphere means that it enjoys powerful synergies with multiple other facilities operating across the electromagnetic spectrum. 

The scientific motivation for LAMBDA is evidenced, for instance, by the 2022 special issue of \textit{Astronomy \& Astrophysics} on low-frequency VLBI with LOFAR \citep[see, e.g.,][]{morabito22} as well as the JUMPING JIVE project \citep{cimo2018}, which advocated for longer-baseline capabilities of SKA-Low. Here, we lay out a number of prominent scientific opportunities.

\subsection{Exoplanets via star-planet interaction}

One of the most exciting prospects is the detection of highly circularly polarized radio emission produced by electron-cyclotron maser instabilities (ECMI), arising either from a star–planet interaction or directly from an exoplanet. Because ECMI emission scales with the magnetic field strength, it offers a direct means to measure the magnetic field of the emitting body. For both star–planet interactions and the strongest magnetic fields predicted by Jupiter-mass dynamo models \citep{callingham24}, this emission can extend to frequencies accessible to SKA-Low \citep{Kavanagh01.2026.SKA}. Even under the most favourable conditions, detecting such signals requires the sensitivity of SKA-Low, particularly for emission originating directly from an exoplanet.

While SKA-Low will be crucial for identifying promising candidates, LAMBDA could provide decisive confirmation by pinpointing the exact location of the emission, distinguishing between stellar and planetary origins. With its unique combination of resolution and sensitivity, LAMBDA will be the only telescope capable of achieving this science case. LAMBDA could establish radio astronomy not only as a tool for measuring fundamental exoplanetary parameters inaccessible at other wavelengths, but also as a powerful new engine for exoplanet discovery.

\subsection{Resolved mapping of AGN}

One of the dominant populations of sources in the radio sky are active galactic nuclei (AGN). These objects form when a supermassive black hole hosted at the center of a galaxy accretes its surrounding gas and dust. During this accretion process, a high fraction of the rest mass energy of this matter is released in the form of electromagnetic radiation and relativistic jets. This form of feedback has a strong effect on the environment of the AGN and is a key process in the formation and evolution of galaxies \citep{harrison24}. In particular the relativistic jets can propagate away from the AGN and form large radio lobes over hundreds of kiloparsecs, enabling the AGN to not only affect their host galaxies, but even the wider environment of those galaxies \citep[see, e.g.,][]{Baczko01.2026.SKA}.

Augmented with LAMBDA, the SKA-Low will be able to map the relativistic jets at (sub-)kiloparsec scales, providing an unprecedentedly detailed view of the jets. Such a spatial resolution is key to study compact radio sources like Compact Symmetric Objects (CSOs) and Compact Steep Spectrum (CSS) sources \citep{odea20}. Furthermore, high spatial resolutions are key to performing polarization measurements, which otherwise are prone to suffer from beam depolarization caused by small-scale magnetic field fluctuations \citep[e.g.,][]{sebokolodi21}. The combination of sensitivity provided by the SKA-Low and the angular resolution enabled by LAMBDA also enhance the study of AGN jets and lobes in the high-redshift Universe, for instance in galaxy proto-cluster environments \citep{overzier16}. Finally, the recent discovery of narrow synchrotron threads associated with AGN jets and lobes \citep{ramatsoku20,rudnick22} also forms a unique science case where (sub-)kiloparsec scale spatial resolutions at low radio frequencies are essential.

As the relativistic jets propagate away from their AGN, energy losses will cause the highest-energy cosmic rays within these jets to rapidly delay to lower energies. This process can be probed by the steepening of the spectrum of the synchrotron emission produced by the relativistic jets, and spectral modelling can be used to age the radio-emitting plasma \citep{harwood13,harwood15}. Through spectral analysis, it is possible to trace the evolution of the relativistic jets and constrain both how these affect their environment and vice versa. At frequencies above 3~GHz, the SKA-Mid will be able to probe sub-kiloparsec scales at any redshift. With LAMBDA, the SKA-Low can match this angular resolution at frequencies as low as 100~MHz, providing the unique opportunity to perform detailed spectral analysis across the majority of the radio window. Spatially resolved information is crucial for providing physically meaningful spectral analysis \citep{harwood17}. 

\subsection{Young stellar objects and supernova remnants}

Within our own Galaxy, radio observations have provided a key window into the birth and death of stars. During the formation of a Young Stellar Object (YSO), the accretion of surrounding matter leads to the launching of jets and outflows, which enable angular momentum to escape from the accretion disk \citep[see, e.g.,][]{anglada18,Sabatini01.2026.SKA}. The jets produced during star-formation are therefore a key component of the formation of young stars. The jets and outflows are typically observed at centimeter frequencies where free-free emission dominates the spectrum, but there is particular value in lower-frequency observations as the turn-over frequency allows the conditions in the ionized plasma to be constrained \citep{ainsworth16}. Additionally, the detection of synchrotron radiation in a proto-star \citep{carrasco10}, despite the jet velocities being substantially lower than in AGNs, further motivates observations in the low-frequency regime.

Higher-resolution observations at low frequencies are critical to the study of jets and outflows associated with YSOs. With LAMBDA, the SKA-Low will be able to probe scales of the magnetosphere (\(\sim\)10~au) at distances of up to 100~pc, enabling the turn-over frequency in the jet-launching region to be mapped. Similarly, LAMBDA will be able to resolve the synchrotron jets in unprecedented detail as they propagate away from the YSO. The collimated inner jet could be mapped in systems out to a distance of 1~kpc from Earth while the larger parsec-scale flows will be resolved in detail throughout our Galaxy.

At the end of a massive star's lifetime, low-frequency radio observations are again key to study the events during and after a supernova. Throughout the lifetime of a supernova remnant (SNR), a central pulsar can accelerate particles up to highly relativistic energies. Furthermore, as the shock wave of the supernova expands into the surrounding interstellar medium (ISM), particles can also be accelerated in-situ through mechanisms such as diffusive shock acceleration \citep[DSA;][]{fermi49}. These accelerated particles can be detected through their synchrotron emission, and provide a substantial contribution to the energy density of the ISM. Low-frequency radio observations are key to disentangle the non-thermal emission from the bright free-free emission \citep[e.g.,][]{tsalapatas24}, allowing particle acceleration theories to be tested. Additionally, low-frequency observations are ideal to detect SNRs long after the initial event \citep[e.g.,][]{arias22}.

\subsection{Pulsar astrometry and scintillometry}

Pulsars have been one of the cornerstones of radio astronomical science since their discovery. Following a core-collapse supernova of a star with a mass between \(\sim\)\(8-20\ \mathrm{M}_\odot\) \citep{smartt09}, the stellar core may be compressed into a neutron star. Depending on the magnetic dipole and the orientation with respect to Earth, such a neutron star may display periodic bursts of radio emission. These objects are known as pulsars, which are unique laboratories for the extreme physical conditions and exotic forms of matter present in these environments.

One of the aims of the SKAO is to exploit its wide field-of-view, multi-beam capabilities and next-generation sensitivity to reveal a large population of pulsars \citep{Keane01.2026.SKA}. Augmented with LAMBDA, the SKA-Low will be able to not only find faint and distant pulsars, but also determine their astrometry with \(\sim\)\(0.1^{\prime\prime}\) precision. Such astrometry can be key to constrain the host environment of a pulsar.

Furthermore, long-baseline observations are also increasingly desirable for the purpose of scintillometry. By observing the difference in the propagation of the radio waves along two different sight lines, the structure of the ISM can be carefully characterized. LAMBDA, with its megametre-scale baselines, will be the only instrument capable of sampling these scales at such low frequencies \citep{Chhetri01.2026.SKA}.

\subsection{Gravitational lensing}

The SKA-Low is set to perform one of the deepest radio sky surveys ever performed, resulting in the detection of millions of radio sources for the first time. With this, a great opportunity is created for the discovery of new gravitational lenses. As the light emitted by a background source passes by a massive structure on its way to Earth, the path that this light follows is bent. Depending on the mass and the geometry of the object along the line of sight, this can result in the background source being multiply imaged \citep[e.g.,][]{saha24}. Such gravitational lens systems a unique cosmological probes as they directly trace the total mass distribution of the lens, as well as the geometry of space \citep{Pandey-Pommier02.2026.SKA}.

High-resolution radio observations are key to study gravitational lenses due to the typical angular separation between multiply-imaged systems. While a galaxy cluster can produce image separations up to arcminute scales \citep[e.g.,][]{heywood21}, most galaxies will produce image separations of only arcsecond scales, or less \citep[e.g.,][]{badole22,jackson24}. With LAMBDA, the SKA-Low will not only be able to detect these lenses, but also have the angular resolution to resolve the separate images of a strong gravitational lens and distinguish these images from any potential foreground emission produced by the lens itself. Finally, at low radio frequencies, LAMBDA also has the unique opportunity to detect plasma lensing \citep{er14}.

\subsection{Fast radio bursts}

The detection of brief (\(\sim\)millisecond) flashes of radio emission \citep{lorimer07}, called Fast Radio Bursts (FRBs) sparked one of the most active fields of radio astronomy \citep[e.g.,][]{lorimer24}. These FRBs remain of largely unknown physical origin and are one of the main open mysteries that the SKA will be trying to solve \citep{Curtin01.2026.SKA}. The dispersion measures of the FRBs reveal that they are typically of extragalactic origin, making them not only interesting phenomena in their own right, but also unique ways to trace propagation properties through the Universe \citep{petroff19}.

Accurately determining the origin of an FRB event is key to both understanding their physical properties and the line of sight that governs the propagation of the signal. The SKA-Low is a particularly important instrument for the study of FRBs, as it has both a wider field of view than SKA-Mid \citep{braun2019anticipatedperformancesquarekilometre} and the propagation effects experienced by the FRB signal also tend to become stronger towards lower radio frequencies \citep{pilia21}. Combined with LAMBDA, SKA-Low will be able to not only detect and study a large population of FRBs, but also precisely locate their host galaxies.

\subsection{HI Absorption in high-redshift galaxies}

At redshifts above 3, the 21 cm line produced by neutral hydrogen (HI) is redshifted down into the SKA-Low band. This enables the SKA-Low to characterize large numbers of high-redshift galaxies through HI absorption in their cold hydrogen reservoirs \citep{Mahony01.2026.SKA}. The distribution and kinematics of this cold gas is essential to understanding the role that it plays in triggering and sustaining AGN activity and star formation across cosmic time. To accurately map HI absorption through a galaxy requires sub-kpc scale resolution, which LAMBDA is able to provide at such redshifts. Additionally, background AGN can also illuminate the cold hydrogen gas in intervening galaxies, allowing the cold gas distribution to be constrained in those galaxies as well. 

\subsection{Technosignature searches}

The large separation between LAMBDA stations is uniquely valuable for technosignature searches \citep[aka Search for Extra-Terrestrial Intelligence, or SETI;][]{Tremblay01.2026.SKA}. A problem with compact interferometers is that it is difficult to verify whether candidate signals are terrestrial or extraterrestrial in nature. LAMBDA stations will be spaced sufficiently far apart that the RFI they detect will largely be unique to each station and heavily mitigated by the correlation process. Furthermore, any RFI produced by satellites will be suppressed by fringe rotation and the phase characteristics will allow us to differentiate between near-field RFI sources and far-field RFI sources (i.e. technosignatures) as has been demonstrated for signals from the Voyager\,2 spacecraft \citep{kim2025}. The computational power and flexibility of the LAMBDA backend could also be used to trigger appropriate voltage capture based on single-station detections for later VLBI follow-up. This will allow VLBI technosignature searches over much larger fields of view than permitted by conventional VLBI station beamforming. Therefore, LAMBDA is an ideal tool to search for extraterrestrial technosignatures. For any technosignature detected, LAMBDA will be able to precisely locate the origin of the signal through the high angular resolution offered by the array. This is key to enabling rapid follow-up observations of any detection.

\section{Unique science targets}

Access to the Southern hemisphere allows LAMBDA in combination with SKA-Low to study several high-profile sources which are poorly accessible from the Northern hemisphere, if at all. Examples of AGN include Centaurus\,A, which is the nearest powerful AGN to us, making it one of the prime laboratories for jet-launching and particle acceleration physics \citep[e.g.,][]{hess18,janssen21}. Furthermore, Pictor\,A is another valuable target, known for its powerful AGN jet \citep[e.g.,][]{wilson01,hardcastle16}, where LAMBDA will be able to complement a series of highly active multifrequency campaigns \citep[e.g.,][]{andati24,marsango24,shaik24,tugliani25}.

For galactic sources, the Southern hemisphere is richly endowed with the Galactic Center as well as both Magellanic Clouds.
The Galactic Center is naturally the environment within which we can find Sagittarius A* \citep{nord04} as well as notably a collection of narrow non-thermal radio filaments \citep{heywood22}.
Meanwhile, the Magellanic Clouds form excellent opportunities to study star-formation \citep[e.g.,][]{for18} and supernovae \citep[e.g.,][]{sasaki25}.

The Southern hemisphere is also the domain containing the Vela pulsar, one of the brightest and closest pulsars to Earth \citep[][and references therein]{rudak18}. The Vela pulsar is a prototype for pulsar glitches \citep{zhou22} as well as for pulsar jets and pulsar-wind nebulae \citep{pavlov03}, making it a high-profile target for high-resolution radio observations such as those accessible through LAMBDA.

\section{The International LOFAR Telescope as a LAMBDA pathfinder}

In the Northern hemisphere, LOFAR serves as an excellent pathfinder for LAMBDA. Not only has LOFAR proven that low-frequency VLBI is possible, but it also identified the challenges involved in performing high-resolution interferometric observations at low radio frequencies, and largely already overcame these challenges. LOFAR has been performing extensive imaging of radio sources at similar frequencies to SKA-Low. The primary survey performed by LOFAR, the LOFAR Two-Metre Sky Survey \citep[LoTSS;][]{shimwell17,shimwell19,shimwell22} has imaged most of the Northern hemisphere at frequencies between 120 and 168~MHz with an angular resolution of 6~arcseconds. Additionally, although not yet a standard operating mode, higher resolutions are achievable through LOFAR's international stations \citep{morabito22}.

One of the main lessons learned from the International LOFAR Telescope (ILT) is to what extent the ionosphere affects our ability to calibrate long-baseline data, and how we can best perform this calibration. The international LOFAR stations heavily lean on a dense catalog of bright and compact radio sources to act as dispersive delay calibrators. For an angular resolution of \(\sim\)0.3~arcseconds, LOFAR has approximately one delay calibrator per square degree available \citep{jackson16,jackson22}. This is also the density required to match the typical isoplanatic patch size of the ionosphere, of approximately one degree in scale. Using the typical properties of LOFAR's delay calibrators, LAMBDA will be able to start a directed search in the Southern hemisphere to search for similar sources and enable the calibration for ionospheric perturbations.

The ILT is also an excellent testbed for other aspects of the LAMBDA data processing. For example, interference produced by one of the five brightest radio sources in the sky (the ``A-team'' sources) is a major issue for low-frequency phased arrays due to their wide field of view and often difficult side-lobe structure. The ILT has been able to observe targets at distances of \(20-30^\circ\) from one of these bright sources without trouble, and with its similar field of view, LAMBDA will likely see a comparable impact from A-team sources. VLBI with the ILT has also had to deal with the density of radio sources in the low-frequency, which causes there to be thousands of sources in the field of view at any given time. This is substantially different from high-frequency VLBI, which typically only concerns one source at a time. With the ILT, this causes challenges with isolating individual sources for the purpose of self-calibration. Using a combination of applying \textit{uv}-limits, averaging down the data to what is still feasible given the typical timescale of ionospheric perturbations, and phasing-up groups of stations to narrow down the field of view, the ILT has been able to achieve the desired isolation.

\section{Synergies with Southern observatories}

In addition to the high-profile scientific targets contained in the Southern hemisphere, one of the key strength of LAMBDA that will allow it to go beyond what LOFAR has achieved in the Northern hemisphere is powerful joint opportunities with other observatories in the Southern hemisphere.

The SKA-Mid in South Africa is an obvious counterpart to the SKA-Low and LAMBDA, enabling spectral indices to be constrained across a broad range of frequencies. Thanks to their well-matched angular resolutions, imaging with these two arrays can provide resolved spectral index maps at close to the resolution limit of either array, depending on the exact frequency range considered. In the radio regime, LAMBDA will also provide unique opportunities to follow up sources detected by the MWA or the Australian Square Kilometre Array Pathfinder \citep[ASKAP;][]{hotan21}. For instance, LAMBDA would be able to respond to a trigger from these telescopes to provide precise positioning for FRB events, or provide complementary frequency coverage.

Additionally, the Southern hemisphere also provides synergies with key optical observatories, such as the Vera C. Rubin Observatory \citep{ivezic19} and the European Extremely Large Telescope \citep[E-ELT;][]{gilmozzi07}. The Vera C. Rubin Observatory currently only reaches just into the Northern hemisphere, with most of its sky coverage being contained within the Southern hemisphere. Similarly, the upcoming E-ELT will also primarily observe at negative Declinations. LAMBDA is ideally positioned to follow-up targets observed by these instruments, or vice versa.

In the Southern hemisphere, LAMBDA would also enjoy a well-matched sky coverage to the Atacama Large Millimeter Array \citep[ALMA;][]{wootten09}, which is one of the premier sub-millimeter observatories. ALMA has been particularly productive with active galactic nuclei, young stellar objects and gravitational lenses. Thanks to its comparable angular resolution to LAMBDA, ALMA forms an excellent higher-frequency counterpart \citep{Forbrich01.2026.SKA}.

Finally, Sunyaev-Zel'dovich (SZ) surveys are also primarily confined to the Southern Hemisphere, with the South Pole Telescope \citep[SPT;][]{ruhl04}, the Atacama Cosmology Telescope \citep[ACT;][]{choi18} and the Simons Observatory \citep{ade19} all being located in the Southern hemisphere. Primarily for the study of galaxy clusters, this makes LAMBDA an ideal instrument to follow-up SZ detections and obtain high-resolution radio observations at low frequencies \citep{Perrott01.2026.SKA}.

\section{Summary and conclusions}

The Square Kilometre Array Observatory is set to become the premier radio observatory of the 2030's thanks to the unprecedented sensitivity of its SKA-Low and SKA-Mid components. However, while the SKA-Low will be world-class in terms of sensitivity, its angular resolution is not as competitive as the SKA-Mid's or the ILT in the Northern hemisphere. In this chapter, we have outlined the plan and goals for the Low-Frequency Australian Megametre Baseline Demonstrator Array (LAMBDA). By distributing 4-6 stations based on the SKA-Low design throughout Australia, we can extend the SKA-Low's baselines from 73.4~km to approximately 4000~km. Using these stations, LAMBDA will be able to probe the scientific potential accessible to the SKA-Low at angular scales down to \(\sim\)0.1~arcseconds as well as prove the feasibility of performing low-frequency radio observations at such high angular resolutions, with the end-goal of providing motivation for a future expansion of the SKA-Low array.

\bibliographystyle{abbrvnat-maxbibnames4}
\bibliography{chapter} 

\end{document}